# Investigation of smooth wave fronts using SLM-based phase retrieval and a phase diffuser

M. Agour[1,2], P. F. Almoro[3], C. v. Kopylow[1] and C. Falldorf[1]

[1] Bremer Institut für Angewandte Strahltechnik, Klagenfurter Strasse 2, 28359 Bremen, Germany
[2] Physics Department, Aswan Faculty of Science, South Valley University, 81528 Aswan, Egypt
[3] National Institute of Physics, University of the Philippines, Diliman, Quezon City, 1101 Philippines

email: agour@bias.de

## Summary

A phase retrieval technique for the determination of smooth wave fronts is demonstrated. It is based on a spatial light modulator in the Fourier domain of a 4f-setup which enables rapid measurements and a diffuse illumination of the test object introducing significant diversity. Optical testing of a lens is given as an application.

## Introduction

Phase retrieval is an approach to determine the lateral phase distribution of a wave field from a set of intensity measurements without the requirement of a reference wave. Recovering the phase from intensity measurements can be described as an inverse problem which is mostly solved iteratively. An important requirement to successful phase retrieval is diversity (differences) in the captured intensities [1]. From earlier techniques that use two intensities [1], recent trends favour the use of multiple intensities. They may be captured through mechanical translation of an aperture [2], or by mechanical displacement of the camera along the optical axis [3,4]. The benefit of using additional constraints embodied by the multiple intensity measurements is the enforcement of the inverse problem to converge faster and more precisely. Mechanical sampling of the multiple intensities, however, is time-consuming and prone to misalignment which affects the phase retrieval [5].

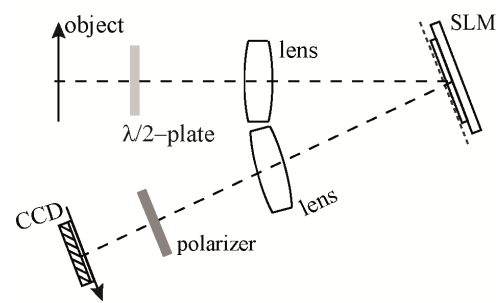

Recently, we introduced the diversity to the investigated wave field by means of a phase-only spatial light modulator (SLM) placed at the Fourier domain of a 4f-imaging system [5,6]. This technique is referred to as SLM-based phase retrieval (SLM-PR) technique. Within the SLM-PR the SLM modifies electronically the wave field by means of the transfer function of free space propagation, which is a pure phase function. Accordingly, a set of intensity measurements associated to different propagated states can be captured across a common camera plane [5]. The SLM-PR setup offers a fast, compact and precise solution to capture the intensities of the wave field associated with different propagated states without the need for any mechanical parts and movements. The effectiveness of the SLM-PR was demonstrated by investigating the deformation of a resistor's surface [7]. Investigation of smooth wave fronts, however, becomes problematic because there is no significant diversity between the intensity measurements corresponding to the different propagation states. A known solution to the problem is achieved by diffused illumination with a partially developed speckle field [8].

Fig 1. Sketch of the SLM-PR setup [6]

In this work, the SLM-PR and the speckle field illumination technique are combined in order to investigate smooth wave fronts. Upon modulations from the phase object and the SLM, the transmitted speckle field which carries the object information varies significantly facilitating the desired diversity. The main advantages of the new SLM-PR technique are fast recording and reconstruction of smooth wave fronts and the automated alignment of the setup. As experimental demonstration, the technique is applied in the shape measurement of thin lenses.

**Experimental results and discussion**

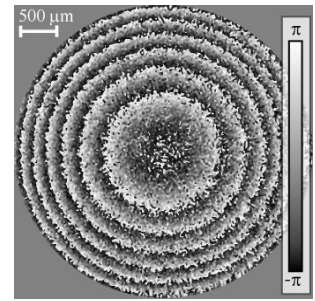

Fig 2. Retrieved phase

Figure 2 shows one example of the retrieved phase of a smooth wave front using the proposed technique. First, a partially-developed speckle field is formed with the aid of a phase diffuser (tuned at 532 nm) which introduces a controlled randomization to an incident plane wave [8]. The speckle field is then used to illuminate a centred thin lens positioned at the object plane in the setup (Fig. 1). The diffuser is 2 cm away from the object plane. A sequence of 10 spatially separated recording planes is selected using the SLM-PR setup, where the distance between successive planes is 3 mm. Subsequently, the speckle measurements with sufficient intensity variation are subjected to an iterative algorithm based on the wave propagation equation to recover the wave field. Figure 2 shows a representative phase distribution at the image plane obtained after 100 iterations. The concentric circular fringes indicate the refractive profile of the lens which is in good agreement with the distribution expected from the known focal length of f=150mm. Additionally, compared to existing techniques involving time-consuming mechanical adjustments, the duration of the measurement process is considerably reduced because the switching time of the employed SLM is only about 50 ms.

**Conclusion**

We have presented an experimental technique for the recovery of a smooth wave front by means of a spatial light modulator and a phase diffuser in conjunction with an iterative phase retrieval algorithm. The application in the shape measurement of a thin lens demonstrates the effectiveness of our technique.